# Electric Field Induced Patterns And Their Temperature Dependence in A Bent-Core Liquid Crystal

Ying Xiang,[1,*] Yi-Kun Liu,[2] Ágnes Buka,[3] Nándor Éber,[3] Zhi-Yong Zhang,[4] Ming-Ya Xu,[1] Everett Wang[1]

*[1]School of Information Engineering, Guangdong University of Technology, Guangzhou 510006, People's Republic of China;*

*(* Author to whom correspondence should be addressed; E-mail: xiangy@gdut.edu.cn ;Frank_xiang68@qq.com)*

*[2]State Key Laboratory of Optoelectronic Materials and Technologies, Sun Yat-sen University, Guangzhou 510275, People's Republic of China*

*[3]Institute for Solid State Physics and Optics, Wigner Research Centre for Physics, Hungarian Academy of Sciences, H-1121 Budapest, Konkoly Thege Miklós út 29-33, Hungary*

*[4]Department of chemistry and Environmental Engineering, Wuhan Polytechnic University, Wuhan 430023, People's Republic of China*

Two kinds of electroconvection (EC) patterns in an ether-bridged bent-core nematic liquid crystal material (BCN), which appear in different frequency ranges, are examined and compared in this paper. One is a longitudinal pattern with the stripes parallel to the orientation of the BCN and with a periodicity of approximately the cell thickness, occurring in the high frequency range of several hundreds Hz; the other one is oblique stripes, which results in a zig-zag pattern, and appears in the low frequency range of several tens Hz. In addition, within an intermediate frequency range, transformations from oblique to longitudinal and then to normal stripes occur at increased ac voltages. In particular, we investigated the temperature behavior of longitudinal and oblique stripes: when the temperature $T$ increases and approaches the clearing temperature $T_c$, the contrast of the domains is enhanced and the frequency range of existence becomes wider, while the onset voltages increase only moderately instead of diverging, thus suggesting an isotropic mechanism of pattern formation.

## 1. Introduction

Liquid crystals (LCs) are one of the best-known classes of functional materials. Being a fluid and ordered medium, LCs possess remarkable electro-optic effects due to their strong anisotropy and sensitive response to external fields, which have widely been used in applications.[1] When driven by a voltage, LCs can undergo different kinds of electroconvection (EC) which arises from the couplings among the deformation of the LC orientation, the flow of both material and charge, and the induced electric field.[2-3] EC phenomena are distinctive features of LCs as nonlinear systems. They often manifest in various patterns. The characteristics of the induced patterns, such as the onset voltage, the critical wave vector and the frequency range of existence, are closely related to the electric transport properties of LCs and the boundary conditions.

Over the past few decades, EC phenomena have been thoroughly investigated for LC systems formed by rod-like molecules.[4-25] The most encountered EC pattern in rod-like systems is normal rolls,[4] which usually occurs in nematics with negative dielectric permittivity and positive electrical conductivity anisotropies ($\varepsilon_a < 0$, $\sigma_a > 0$) in ac fields of a broad frequency range ($f$ ~ 20–1000Hz) above a certain threshold voltage $V_c$ and is characterized by stripes perpendicular to the initial LC orientation $\mathbf{n}_0$ with a periodicity of $\Lambda \approx 2\ d$ ($d$ is the thickness of the LC layer). At first, based on inseminating ideas of Carr [5] and Helfrich [6] a one-dimensional model [7] was constructed to explain the formation of normal rolls via a feedback mechanism: a periodic fluctuation in the director tilt leads to space charge separation due to the anisotropic conductivity, the Coulomb-force on the space charge induces a vortex flow, which exerts a destabilizing torque on the director. This feedback becomes positive at voltages above $V_c$, thus the fluctuations grow to a macroscopic pattern. This theory was later generalized to a three-dimensional, so called standard model (SM) of EC [2,8] which is able to describe besides the normal rolls also the oblique ones, [9] which are characterized by their wave vector forming a non-zero angle with $\mathbf{n}_0$.[10-11] The onset voltage $V_c$ and the critical wave vector $\mathbf{q}_\mathbf{c}$ provided by the SM depend on a huge set of material parameters (permittivities, conductivities, elastic moduli, viscosities and flexoelectric coefficients).[26] The SM was found suitable to explain pattern formation also for nematics with $\varepsilon_a > 0$, $\sigma_a < 0$.[27] By now, all patterns which can be described well by the SM are known as “standard EC”.

There exist, however, so called non-standard EC patterns occurring under conditions for which the SM itself does not predict any instability.[3] One representative is the pattern with longitudinal rolls, i.e. with rolls running parallel or slightly oblique to $\mathbf{n}_0$, observable in nematics with ($\varepsilon_a < 0$, $\sigma_a < 0$).[15-16] The change of the sign of $\sigma_a$ compared to the case discussed above inverts the polarity of the space charges and therefore the direction of the flow; thus the flow induced torque is stabilizing. It has recently been proven [25] that the inclusion of flexoelectricity into the SM provides an interpretation; namely in this case there is an additional charge separation originating from the flexoelectric polarization of the periodically deformed state which drives the convection.[28]

The prewavy or wide domain instability is another kind of non-standard patterns, typically occurring at high $f$ (above a few kHz) in several nematics, and forming stripes normal to $\mathbf{n}_0$ with a periodicity of multiples of $d$ ,[23-24] which are characterized with azimuthal (in-plane) modulation of the director and hence are best viewed between crossed polarizers. Prewavy patterns were observed in nematics with $\varepsilon_a < 0$, $\sigma_a > 0$, as well as in those with $\varepsilon_a < 0$, $\sigma_a < 0$; thus the sign of the conductivity anisotropy seems to be irrelevant. It is unclear yet, what other phenomena should be added to the SM to provide an explanation.

In addition, in some nematics at dc or at very low frequencies ($f$ ~ 0–20 Hz), static longitudinal domains without convection have also been observed.[29-30] This instability (flexoelectric domains) is driven by the energy gain in the periodically deformed state due to the linear coupling between the flexoelectric polarization $\mathbf{P}_\mathbf{fl}$ and the external electric field $\mathbf{E}$ .[31-32]

The pattern morphologies introduced above are common in one respect: the experimentally found (as well as the theoretically predicted) threshold voltages increase while the pattern wavelengths decrease with increasing frequency.

Recently, new kinds of LC systems based on bent-core molecules have been invented and generated considerable interest since their banana-shaped molecular structure leads to peculiar physical properties.[33] Bent-core nematic (BCN) systems (including mixtures of bent-core and rod-like molecules) may also exhibit EC patterns.[34-45] The patterns reported so far can mostly be classified as non-standard EC, however, with some characteristics substantially different from those in calamitic LCs.

The pioneering work [34] reported longitudinal stripes at very low $f$ and two prewavy (PW) regimes at higher $f$ in the BCN ClPbis10BB. The latter possessed threshold voltages diverging as $f$ approached the edges of the frequency gap separating the two prewavy regimes. The high frequency PW regime had some remarkable features: First, its range of existence extended to unusually large frequencies ($f$ > 500 kHz); second, the threshold voltage decreased, unprecedentedly, on increasing $f$. Similar characteristics were later reported for mixtures [38] and for some other BCNs [39-40] too. In another BCN system, CNRbis12OBB, also longitudinal (LS) and normal stripes (NS) were observed in two distinct frequency ranges [41]: in the low-frequency range of 150 Hz-550 kHz, LS appeared with stripes parallel to $\mathbf{n}_0$ and $2\ d \leq \Lambda \leq 3\ d$; whereas in the high-frequency range of 200 kHz–1 MHz, NS appeared with stripes perpendicular to $\mathbf{n}_0$ and $\Lambda \approx 3\ d$. The periodicities of LS and NS varied with the amplitude and frequency of the applied voltage. Furthermore, another kind of LS, possessing strongly temperature dependent periodicity of less than $d$, was reported by S. Kaur et al., [43]where the LS displayed distinct features in the high and in the low temperature part of the nematic phase; for example, the periodicity was almost constant in the high temperature regime whereas it exhibited strong temperature dependence in the low temperature regime. This temperature dependent behavior was ascribed to a transition between uniaxial and biaxial nematic phases.[44]

Bent-core nematics, due to the combination of their molecular shape with a transverse component of their dipole moment, are ideal candidates for compounds with a large flexoelectric response. Indeed, determination of the flexoelectric coefficient of some BCN materials from the observed flexoelectric domains [45] and flexoelectro-optic effect [46] confirmed this theoretical anticipation. Moreover, it was reported that under certain conditions (e.g. in flexible cells subjected to an oscillatory bend deformation) BCN systems may exhibit a giant flexoelectric response, corresponding to a bend flexoelectric coefficient which is several orders of magnitude greater than that of rod-like compounds [47-49].

In view of the rich morphology of electric field induced patterns and the unconventional flexoelectric properties of BCN systems, the study of EC in such materials is a challenging field with the main question: how the nonlinear coupling of relevant physical parameters results in the formation of varied arrays of static and dynamic patterns. This EC investigation is especially useful when carried out supplementing the studies on Freedericksz transitions that are solely characterized by the dielectric anisotropy and the elastic constants of the system.

In this paper, we investigate the characteristics of oblique, longitudinal and normal stripes in an ether-bridged BCN system, where longitudinal stripes are the predominant feature. We examined the critical temperature behavior of these patterns, and observed that their contrast is enhanced with increased temperature. In addition, transformations between these patterns are realized by changing the amplitude and frequency of ac voltage.

## 2 Experimental arrangement

The chemical structure and phase transition sequence of the ether-bridged BCN material 2,5-di{4-[(4-octylphenyl)–difluoromethoxy]-phenyl}-1,3,4- oxadiazole (8P-CF2O-ODBP) [50] are shown in Figure 1.

Samples of the BCN compound were filled into standard planar sandwich cells, made of two ITO glass substrates with a 6 μm spacer in between. The two ITO glass substrates were coated with a high temperature polyimide ensuring that no temperature-induced changes in the alignment should occur, and then rubbed unidirectionally to obtain planar alignment. The initial LC orientation $\mathbf{n}_0$ is along the x-axis, while the direction of the electric field $\mathbf{E}$ is along the z-axis, perpendicular to the substrates and $\mathbf{n}_0$. The temperature of the samples was controlled using a heating stage with a relative accuracy of ±0.1 °C.

Sinusoidal AC voltages up to ~400 Vpp were applied to the samples. The induced EC patterns were observed in transmission, between crossed polarizers, using a polarizing microscope equipped with a high resolution CCD camera.

Besides polarizing microscopy, an optical diffraction technique was also used to probe the characteristics of EC patterns. The periodic director distortions in the stripes form an optical grating inside the samples, and thus diffract the incident He-Ne laser beam ($\lambda$ = 633 nm). Consequently, the characteristics of the patterns can be revealed by monitoring the behavior of the diffraction fringes. For the measurements a slightly focused, linearly polarized laser beam with a spot size of 100 μm was used. The diffracted intensity was measured by a fast photodetector (DET36A/M from Thorlabs).

## 3 Experimental results

The experiments were carried out at several temperatures across the whole nematic phase range, where oblique, longitudinal and normal stripes could be triggered and transformed into each other by varying the temperature, frequency and voltage. In the following, the characteristics of patterns obtained from polarizing microscopy and the diffraction technique are summarized.

### 3.1 Investigation of stationary patterns by polarizing microscopy

Different pattern morphologies appear at a fixed temperature as the ac frequency is swept through the experimentally studied range of 30 Hz < $f$ < 1100 Hz. Figs. 2(a) and 2(b) exhibit the frequency dependence of the onset voltages of various EC patterns far away and close to the clearing temperature $T_c$, respectively. It is seen that the primary instability may manifest itself in two distinct EC morphologies: oblique stripes (OS) and longitudinal stripes (LS). The specific frequencies separating the $f$ ranges of existence of various structures depend strongly on the temperature, as it is demonstrated more expressly by the morphological diagram in Fig.3.

OS occur at low frequencies, for $f_{OL} < f < f_{OH}$ (the region with hatched lines tilting left and the crosshatched region in Fig.3), at voltages $V > V_{cOS}$. The stripes are running at a ± angle, symmetrical with respect to $\mathbf{n}_0$ as in Fig.4(a), with a periodicity of $\Lambda \approx 9$ μm which slightly decreases with the increase of $f$.

At higher frequencies, for $f_{OH} < f < f_{LH}$ (region with hatched lines tilting right in Fig.3), LS are observed which orient nearly parallel to $\mathbf{n}_0$; their periodicity, $\Lambda \approx 5$ μm, is shorter and approximately equates the thickness (d = 6 μm) of the LC film, as shown in Fig.4(b).

In the low temperature part of the nematic phase ($T - T_c < -20$ °C, Fig.2(a)) only these two primary structures could be detected. Above these temperatures there is a transition frequency range, $f_{LL} < f < f_{OH}$ (crosshatched region in Fig.3), where the primary pattern is still OS with $\Lambda \approx 7$ μm, but LS may appear as a secondary instability upon increasing the voltage above $V_{cLS} > V_{cOS}$. While $f_{OL}$ and $f_{LL}$ are almost independent of temperature, both $f_{OH}$ and $f_{LH}$ shift to higher frequencies for higher temperatures. Therefore the transition frequency range broadens as $T$ increases.

Moreover, at temperatures close enough to $T_c$ an additional morphology, normal stripes (NS) appear as a ternary instability for $V > V_{cNS} > V_{cLS}$. The frequency range of existence of NS, $f_{NL} < f < f_{NH}$ , roughly coincides with that of the secondary LS pattern. NS differ very much from OS and LS in their direction as well as in their periodicity: these stripes run perpendicular to $\mathbf{n}_0$ and are much wider, having $\Lambda \approx 2$–$3\ d$ (see Fig.4(c)). This makes us conclude that NS represent a kind of prewavy pattern.

Thus being close to $T_c$ in the transition frequency range, increasing the voltage three distinctive patterns (oblique, longitudinal and normal stripes) occur subsequently, each of them characterized by distinctive, different orientation, periodicity and threshold. If the ac field is increased further, the BCN system turns into a random, chaotic state; the stripes become distorted and then disappear. This pattern sequence is demonstrated in Figs. 5(a), (c), (e) and (f), respectively. Note that in Fig. 5(e) the NS of high contrast coexist with the finer LS; first the LS (Fig.5(f)) then also the NS disappear when the BCN system turns into chaos.

Upon increasing the voltage defect rich transitional states exist in between the various types of domains, as shown in Figs. 5(b) and (d). These states are not stationary, but are characterized by transient generation, motion and finally annihilation of defects, which is the way how the system of stripes accommodates to the new wave vector corresponding to the new control parameters (the frequency and the applied voltage).

We note that rotating the sample with 90° a change of the interference colours was detected as shown in Figs. 4(d)-(f); the pattern morphology otherwise remained unaltered.

The patterns were also checked using a single polarizer only (shadowgraph image). For OS and LS no shadowgraph image was obtained, while for NS only a very faint pattern could be observed.

We note that when the applied voltage is increased or decreased repeatedly, all pattern types and transition states are re-established without hysteresis.

We mention that if the excitation frequency is below $f_{OL}$ or above $f_{LH}$, no regular stripe pattern could be detected; rather at high voltages only a strongly scattering, chaotic state appears.

The variation of the threshold voltages in Fig. 2 and the morphological diagram of Fig. 3 already indicated that temperature plays an important role in determining the character of pattern formation. In addition, as the temperature increases, patterns become better defined (easier to detect) independent of the actual morphology, as shown by the image sequences in Fig. 6.

Besides their dependence on the frequency, the onset voltages $V_{cOS}$, $V_{cLS}$ and $V_{cNS}$ depend also on $T$, which needs some additional addressing. Typically the onset voltages increase with increasing $T$. In order to test the behavior close to the phase transition which may give some clue about the origin of the pattern forming phenomenon, we carried out the measurements approaching $T_c$ in small temperature steps, going as close as possible. The results shown in Table 1 indicate that the threshold voltages of all pattern types increase moderately and smoothly, without diverging when going toward $T_c$, indicating the possibility of an isotropic pattern forming mechanism.

In addition to threshold determination, the morphological changes of the patterns in the close vicinity of $T_c$ were also tracked. It was found that the patterns remain of high contrast and well observable when $T \rightarrow T_c$. Thermostatting the sample at about the phase transition temperature, $T \approx T_c$, a coexistence of the nematic and the isotropic phases could be found (due to a little temperature inhomogeneity). It is clearly seen in Figs. 7(a) and (b), that the patterns exist in the nematic side of the sample, up to the phase boundary. In the isotropic region no stripes are visible. Note that in the isotropic phase ($T > T_c$) the lack of anisotropy and the resulting loss of birefringence prevents from detecting flow with polarizing microscopy, even if convection were present.

### 3.2 Investigation of the dynamical behavior of the patterns by diffraction technique

A periodic director distortion corresponds to an optical grating with amplitude as well as with phase modulation. Both are directly related to the pattern amplitude $A$ (the maximum director tilt). The relation between the intensities $I_n$ of the $n$th order diffraction fringes and $A$ has recently been calculated rigorously for an arbitrary director distortion profile.[51] In accordance with previous approximations it was found that at normal incidence of light the amplitude grating contributes mostly to the 1st order fringe, while the phase grating determines the 2nd order one. For the small amplitudes occurring in standard EC patterns near onset, at normal incidence $I_2 \propto A^4$ dominates, while $I_1 \propto A^2$ becomes notable mainly at oblique incidence. [52]

Our observations have shown that whether the probe beam is incident normally or slightly obliquely (by about several degrees), the resultant diffraction patterns are almost identical. Namely, both the 1st and 2nd order diffraction fringes are present. This holds for OS, for LS as well as for NS patterns. It was checked that the diffraction angle of 1st order fringe corresponds to the periodicity of the stripes seen in the polarizing microscope.

In order to explore the anisotropic features of the diffraction, we checked the diffraction fringes at the polarization of the laser beam **p** being either parallel or perpendicular to the initial director orientation $\mathbf{n_0}$. In standard EC (normal rolls) diffraction is typically observed only with $\mathbf{p} \parallel \mathbf{n_0}$. In contrast to this, in our case strong diffraction fringes were observable at both polarization states: $\mathbf{p} \parallel \mathbf{n_0}$ and $\mathbf{p} \perp \mathbf{n_0}$. This statement holds for all three pattern morphologies. The presence of diffraction fringes at $\mathbf{p} \perp \mathbf{n_0}$ clearly indicates that all the nonstandard pattern types discussed in this paper involve azimuthal (in-plane) rotation of the director. This agrees with previous polarization microscopic observations on the prewavy pattern (which NS are thought to be) in calamitic LCs [23]. Moreover, it also agrees with the theoretical predictions of the SM extended with flexoelectricity [25] for the longitudinal rolls (which LS are assumed to be).

The polarization states of the transmitted beam (0th order) and the 1st and 2nd order fringes were also determined. It was found that the polarization of the 1st order fringe was perpendicular to **p**, while the polarization for the 2nd order fringe was parallel with **p**, independent of the direction of **p** with respect to $\mathbf{n_0}$. The 0th order fringe had an elliptical polarization, with its long axis almost parallel to **p**. These observations hold for OS and LS as well as for NS.

Monitoring the intensities diffracted on the patterns induced by ac voltages provides information on the temporal evolution of the director distortions. Knowing this is especially important, as for the standard EC the SM predicts two types of solutions (temporal modes): one with a director field stationary (in leading order) within the period of the driving ac voltage (the conductive regime) and another with the director oscillating with $f$, thus changing the sign of the tilt periodically (the dielectric regime) [2,8]. By diffraction these two regimes can easily be distinguished.[20-21] Extending the SM by including flexoelectricity has shown that flexoelectricity mixes the modes [25]: the solutions for both standard EC regimes are combinations of the stationary and the oscillating modes, with one of them dominating over the other. This mixing could, however, not been identified in the diffraction intensities, indicating weak mixing due to the small flexoelectric coefficients.

Applying the diffraction method to the nonstandard EC scenarios similarly provides insight on how the director rearranges itself upon ac driving. Fig. 8 shows the temporal dependence of the intensity $I_1$ of the 1st order diffraction fringe of various patterns measured with a fast photodetector and compares to that of the applied voltage. It is seen that when the frequency is high ($f$ = 500 Hz), the diffraction intensities remain roughly constant as time evolves (see Figs. 8(b), (c) and (d) for oblique, longitudinal and normal stripes, respectively), which means that the director is stationary in the leading order, the small $2f$ modulation in the intensity comes from the quadratic term. However, if the frequency is low ($f$ = 60 Hz), the diffraction intensity oscillates strongly with a frequency of $2f$, almost following the $I_1 \propto \sin^2(2\pi f\, t)$ dependence (see Fig. 8(a) for oblique stripes), suggesting that the director oscillates with the driving frequency. Note that the change of the polarization direction of the illuminating laser beam does not change the temporal characteristics.

So far a stationary (in leading order) director field was detected in the conductive regime of standard EC (in accordance with the theoretical expectations) as well as in the prewavy pattern of compounds exhibiting standard EC. As the NS pattern in 8P-$CF_2$O-ODBP is assumed to be a kind of PW, the near constancy of the intensity diffracted on NS (Fig. 8(d)) is not surprising. In contrast to that, the stationary intensity diffracted on LS and OS at high $f$ (Figs. 8(b) and (c)) is quite unexpected, as earlier experiments [21] on the nonstandard longitudinal EC patterns of calamitic nematics indicated director oscillations through the initial state (100% intensity modulation), similar to those in the standard dielectric regime.

The scenario, that a high $f$ stationary pattern transforms into a seriously time modulated one, is not unprecedented. Recently this has been demonstrated for the conductive regime of standard EC [53] where the transition takes place in the frequency range corresponding to the inverse director relaxation time. The frequency of 60 Hz in Fig. 8(a) seems, however, to be too large for the same interpretation to hold. As the viscosities and consequently the director relaxation time of BCN compounds are usually higher than those of calamitic nematics, the inverse relaxation time of 8P-$CF_2$O-ODBP is expected to be much below 60 Hz.

We note that a large (30–50%) modulation of the diffracted intensity was reported for travelling waves in the conductive regime of standard EC rolls as well as for nonstandard travelling longitudinal rolls [21]. The patterns discussed in this paper, however, do not travel.

The occurrence of longitudinal stripes is thought to be the consequence of the flexoelectric properties of the studied compound [25]. Due to its bent molecular shape, 8P-$CF_2$O-ODBP is expected to have larger flexoelectric coefficient than calamitic nematics. Therefore the flexoelectricity induced superposition of modes with stationary and oscillating director deformations may be more pronounced, which might be a clue for the observed large intensity modulation.

### 3.3 Temperature and frequency dependences of the dielectric permittivity and electrical conductivity of the BCN material

As the anisotropies of the electric parameters (dielectric permittivity and electric conductivity) are key factors determining the behavior of EC patterns, we carried out measurements of these parameters.

We note that the molecular structure as well as the absence of uniform Freedericksz transition in the planar cell at all considered temperatures and frequencies implies that the BCN material should exhibit negative dielectric anisotropy $\Delta\varepsilon = \varepsilon_{||} - \varepsilon_{\perp} < 0$.

Nevertheless we attempted a direct measurement of the electric anisotropies. A technical problem occurred, however, because the polymers usually applied to obtain homeotropic alignment for rod-like LCs did not provide alignment for the tested BCN LC. Therefore, in order to determine the anisotropy, we had to use two, complementary, in-plane switching (IPS) cells with planar alignment, which were identical in structure (cell thickness of $d$ = 4 μm and $g$ = 40 μm gap between the electrodes) apart from their rubbing directions $\mathbf{n}_0$. One IPS cell, where the directions of $\mathbf{n}_0$ and the electric field **E** coincided, enabled the measurement of the parallel permittivity $\varepsilon_{||}$ and conductivity $\sigma_{||}$; the other IPS cell, with **E** orthogonal to $\mathbf{n}_0$, served to determine the perpendicular permittivity $\varepsilon_{\perp}$ and conductivity $\sigma_{\perp}$.

Clearly, the electric field is not homogeneous in the IPS cell, which reduces the accuracy of the measurement and may provide smaller values for the anisotropies than the actual ones. This can, however, be partially compensated by performing measurements on a well known reference material 5CB (pentyl-cyano-biphenyl) for calibration purposes. Despite of these problems the technique is suitable to find, qualitatively, the main tendencies in the frequency and temperature dependences. The measurement results, which were obtained by an Agilent E4890A Precision Component Analyser, indicate that the BCN material possesses negative dielectric and negative conductive anisotropies. Both anisotropies are relatively small, but their negative signs persist over the entire nematic temperature range for all considered frequencies, similar to that of other BCN materials [42].

## 4 Discussion and Conclusions

The results presented above represent the first experimental tests, including an in-depth exploration of the temperature, frequency and voltage dependence of pattern formation and its dynamics as well as determination of the polarization state of the diffraction fringes and of the signs of the electric anisotropies, for a newly synthesized bent-core nematic compound 8P-$CF_2$O-ODBP.

As in this compound both the dielectric and the conductivity anisotropies are negative, the standard Carr-Helfrich mechanism alone cannot cause destabilization; thus standard EC patterns cannot occur. Consequently the patterns described above are similar to those found in other BCN compounds in the sense that they all represent nonstandard EC scenarios. The studied 8P-$CF_2$O-ODBP, however, behaved differently from other BCNs like ClPbis10BB, as it did not exhibit EC at very high ($f > 10$ kHz) frequencies and, therefore, no EC regime was detected where the onset voltage decreases with increasing frequency. This may be in correlation with the absence of strong dielectric relaxation at low frequencies, which guarantees the constancy of the signs of the dielectric and conductivity anisotropies in the studied frequency range.

On the other hand, the ability to tune the pattern characteristics/morphologies by frequency as well as with voltage or temperature, strongly reminds of the behaviour of the rod-like nematic 8/7 (4-*n*-octyloxy-phenyl-4-*n*-heptyloxy-benzoate) which (being a calamitic compound with negative dielectric and conductivity anisotropies at low temperatures) exhibited to some extent similar features.[16] An important difference is, however, that in 8/7 this tunability occurred at around the temperature of the sign inversion of $\sigma_a$, where a transition from nonstandard EC (longitudinal rolls) to standard EC (normal rolls) through nonstandard and standard oblique rolls could be observed. In contrast to this, our compound preserves the negative sign of $\sigma_a$ in the whole nematic temperature range and all exibited patterns, including the prewavy-like normal stripes, are of nonstandard type. Therefore the tunability of the morphologies should have different origin.

The strongly oblique stripes found at low $f$ (Fig. 4(a)) and the unexpected frequency dependence of their dynamics seem to correspond to a new scenario for nonstandard EC. The occurrence of nonstandard longitudinal rolls in 8/7 could be explained by the SM extended with flexoelectricity [25,32]; the same interpretation is expected to apply to the longitudinal stripes observed in 8P-$CF_2$O-ODBP and might account for the oblique ones too. Numerical analysis of the SM extended with flexoelectricity would be necessary to calculate $V_c(f)$, $\mathbf{q}_c(f)$ and the temporal evolution of the director field, however, such simulations could not yet been performed for the unusual parameter ranges of bent-core systems.

The behavior close to (or at) the nematic-isotropic phase transition temperature $T_c$ generally allows drawing some conclusions on the mechanism of the pattern formation. In standard EC formation of patterns is governed by the Carr-Helfrich mechanism that relies on the anisotropy of the nematic. The BCN material 8P-$CF_2$O-ODBP exhibits, however, nonstandard EC for which it is not unambiguously clear if the driving mechanism is of anisotropic or isotropic origin. Though a rigorous theoretical description of the isotropic mechanism[1,54,55] is not yet available, one may expect that the relevant physical parameters (an average of the anisotropic parameters) change smoothly when the temperature approaches or goes through $T_c$; therefore $V_c$ should not depend strongly on temperature. In contrast to that, the anisotropies of the physical quantities typically exhibit strong temperature dependence when the phase transition is approached (where all anisotropies disappear), which is expected to result in a substantial increase of the onset voltage(s). One should remember, however, that a critical behavior (divergence of $V_c$) is strictly expected for a continuous, second order phase transition only. The nematic to isotropic transition is, however, a (though weakly) first order one, with a (small) jump in the anisotropy values at $T_c$. The strong temperature dependence of the anisotropic material parameters still mostly holds, but to what extent that increases $V_c$ at $T \to T_c$, depends on the compound.

The measurements of the onset voltages of 8P-$CF_2$O-ODBP have shown neither a divergence, nor a substantial increase of $V_c$ when approaching $T_c$. The presence of patterns in the nematic region close to the phase boundary in the two-phase coexistence case also supports the absence of strong $V_c$ increase for $T \to T_c$. This might be an indication that the patterns in 8P-$CF_2$O-ODBP could be induced by an isotropic mechanism. The above observations, however, cannot be regarded as an unquestionable proof for the same.

## Acknowledgements

Authors are indebted to Professor J. Y. Zhou in Sun Yat-sen University for numerous useful discussions. This work was supported by the National Natural Science Foundation of China (11374067, 11074054), the Hungarian Research Fund (OTKA K81250) and the Chinese-Hungarian bilateral project TÉT_12_CN-1-2012-0039.

**Figure and table list**

**FIG. 1** Chemical structure and phase transition sequence of the BCN material 8P-$CF_2O$-ODBP.

**FIG. 2** (Color online) Frequency dependence of the onset voltages $V_c$ associated with oblique (OS), longitudinal (LS) and normal (NS) stripes (a) at the lower temperature range of the nematic phase, (b) at temperatures close to the clearing point.

**FIG. 3** (Color online) Frequency range of existence for electric field induced patterns at different temperatures. Lines hatched to left and right mark the regions where oblique and longitudinal stripes, respectively, appear as the primary instability. In the crosshatched region morphological transitions occur upon increasing the voltage.

**FIG.4** (Color online) Characteristic pattern morphologies observed in a polarizing microscope: (a) and (d) oblique stripes with ≈ 9μm periodicity at $T$-$T_c$= -5ºC, $f$ = 80Hz, $V$ = 40$V_{pp}$; (b) and (e) longitudinal stripes with ≈5 μm periodicity at $T$-$T_c$= -5ºC, $f$ = 500Hz, $V$ = 80$V_{pp}$; (c) and (f) normal stripes with ≈ 17 μm periodicity at $T$-$T_c$= -1ºC, $f$ = 500Hz, $V$ = 160$V_{pp}$. The inscriptions A, P and rubbing in the microphotographs (a) and (d) refer to the analyzer, polarizer and the rubbing direction of the LC cell, respectively. The different colours in morphologies according to different orientation indicate the birefringence of these stripes.

**FIG. 5** (Color online) Voltage dependence of the patterns at the temperature $T$-$T_c$ = -3 °C and frequency $f$ = 500 Hz, illustrated by a sequence of microphotographs: (a) oblique stripes at $V$ = 58 $V_{pp}$; (b) transformation from oblique to longitudinal stripes at $V$ = 62 $V_{pp}$; (c) longitudinal stripes at $V$ = 86 $V_{pp}$; (d) transformation from longitudinal to normal stripes at $V$ = 110 $V_{pp}$; (e) normal stripes (coexisting with the fine longitudinal ones) at $V$ = 146 $V_{pp}$; (f) normal stripes at $V$ = 250 $V_{pp}$ before entering the chaotic regime. The microphotographs cover an area of 50 μm × 50 μm. The arrow in (f) indicates the initial director orientation (rubbing direction).

**FIG. 6** (Color online) The effect of temperature increase on various pattern morphologies: oblique stripes ($f$ = 125 Hz) at (a) $T$ - $T_c$ ≈ -15 ºC, $V$ ≈ 12 $V_{pp}$, (b) $T$ - $T_c$ ≈ -10ºC, $V$ ≈ 32 $V_{pp}$, (c) $T$ - $T_c$ ≈ -5ºC, $V$ ≈ 46 $V_{pp}$; longitudinal stripes ($f$ = 500 Hz) at (d) $T$ - $T_c$ ≈ -15ºC, $V$ ≈ 70 $V_{pp}$. (e) $T$ - $T_c$ ≈ -10 ºC, $V$ ≈80 $V_{pp}$. (f) $T$ - $T_c$ ≈ -5 ºC, $V$ ≈ 84 $V_{pp}$; normal stripes ($f$ = 400 Hz) at (g) $T$ - $T_c$ ≈ -12 ºC, $V$ ≈ 120$V_{pp}$, (h) $T$ - $T_c$ ≈ -10 ºC, $V$ ≈ 120 $V_{pp}$, (i) $T$ - $T_c$ ≈ -5 ºC, $V$ ≈ 120 $V_{pp}$.

**FIG. 7** (Color online) Electroconvection at the clearing temperature ($T$ = $T_c$) . The microphotographs show a coexistence of the nematic (upper right corner) and isotropic (bottom left corner) phases separated by a sharp phase boundary line. EC pattern exists in the nematic part: (a) oblique stripes ($f$ = 150 Hz) at $V$ ≈ 60 $V_{pp;}$ (b) longitudinal stripes ($f$ = 500 Hz) at $V$ ≈ 88 $V_{pp}$

**FIG. 8** (Color online) Temporal evolution of the applied voltage and of the 1$^{st}$ order diffraction intensity at two orthogonal polarization directions of the illuminating light for various patterns under ac driving at the temperature $T$-$T_c$ ≈ -3 ºC: (a) oblique stripes (OS) at $V$ ≈ 35 $V_{pp}$, $f$ = 60 Hz; (b) oblique stripes (OS) at $V$ ≈ 64 $V_{pp}$, $f$ = 500 Hz; (c) longitudinal stripes (LS) at $V$ ≈ 76 $V_{pp}$, $f$ = 500 Hz; (d) normal stripes (NS) at $V$ ≈ 120 $V_{pp}$, $f$ = 500 Hz.

**Table.1** The temperature dependence of the onset voltages $V_c$ at $f$ = 500 Hz in the vicinity of the clearing point $T_c$.

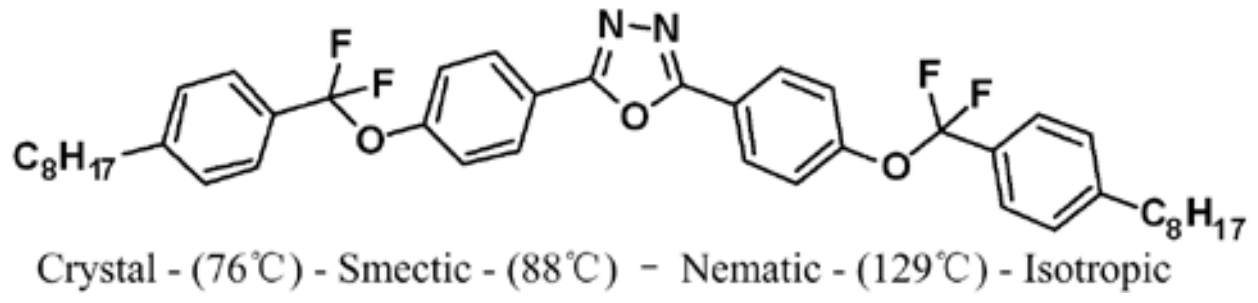

**FIG. 1** Chemical structure and phase transition sequence of the BCN material 8P-$CF_2$O-ODBP.

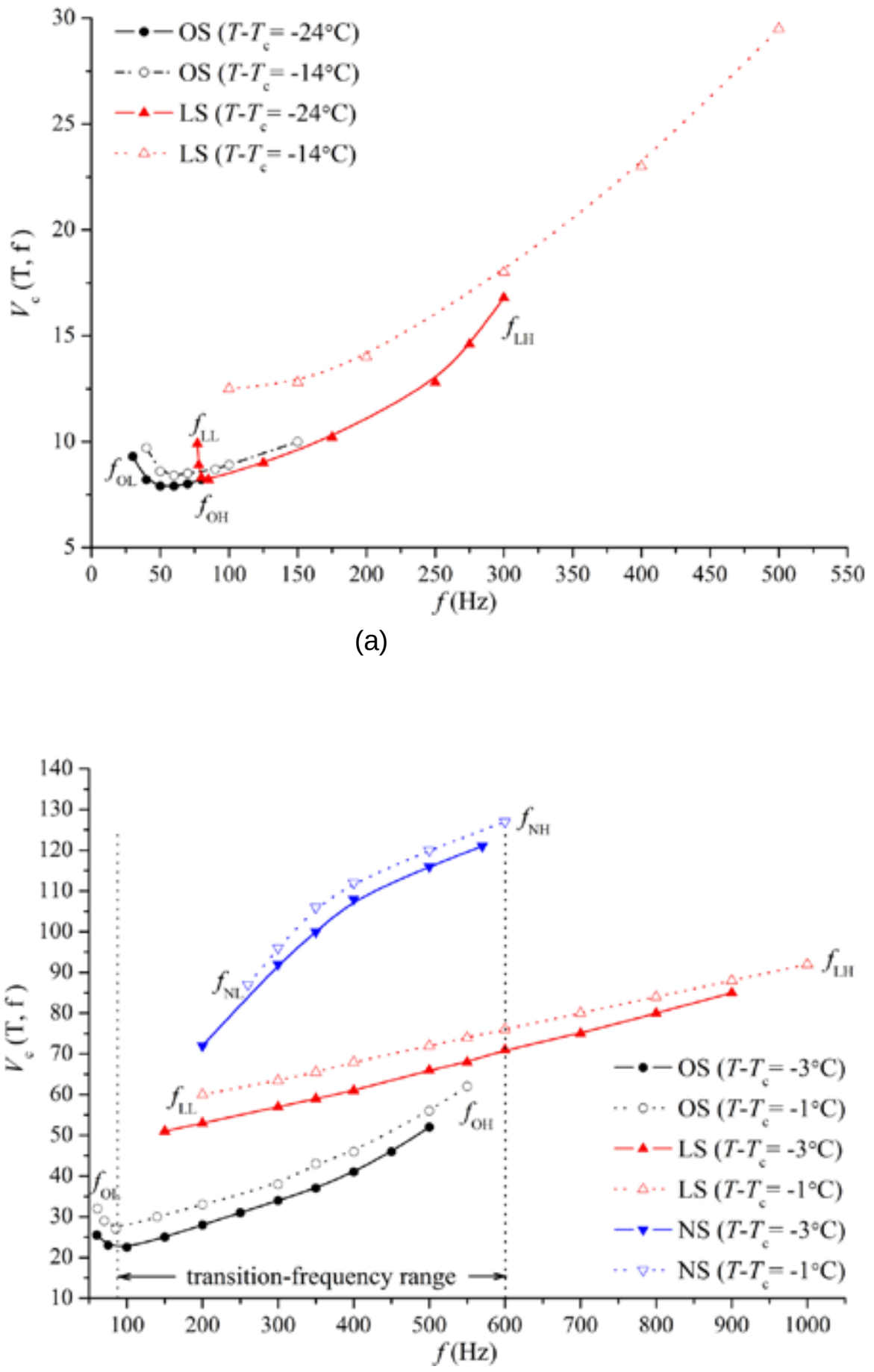

(b)

**FIG. 2** (Color online) Frequency dependence of the onset voltages $V_c$ associated with oblique (OS), longitudinal (LS) and normal (NS) stripes: (a) at the lower temperature range of the nematic phase, (b) at temperatures close to the clearing point.

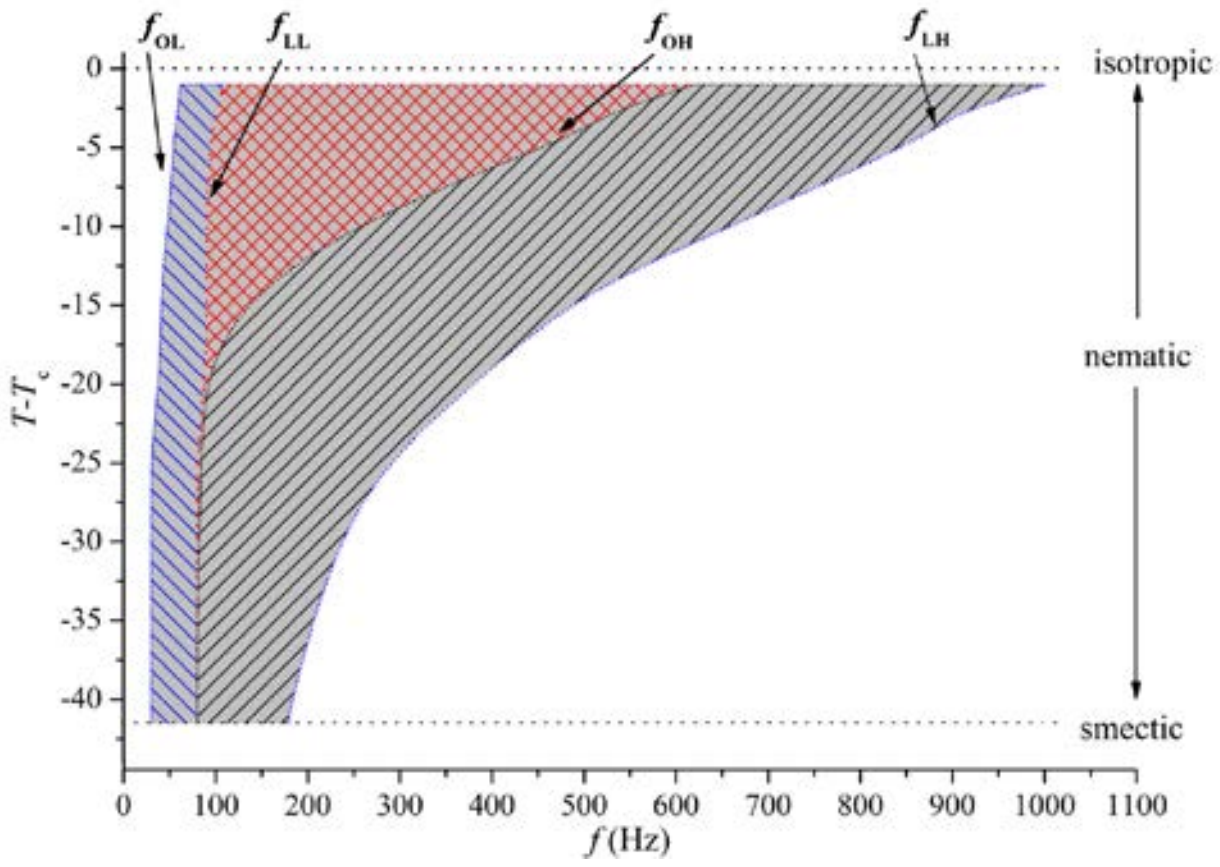

**FIG. 3** (Color online) Frequency range of existence for electric field induced patterns at different temperatures. Lines hatched to left and right mark the regions where oblique and longitudinal stripes, respectively, appear as the primary instability. In the crosshatched region morphological transitions occur upon increasing the voltage.

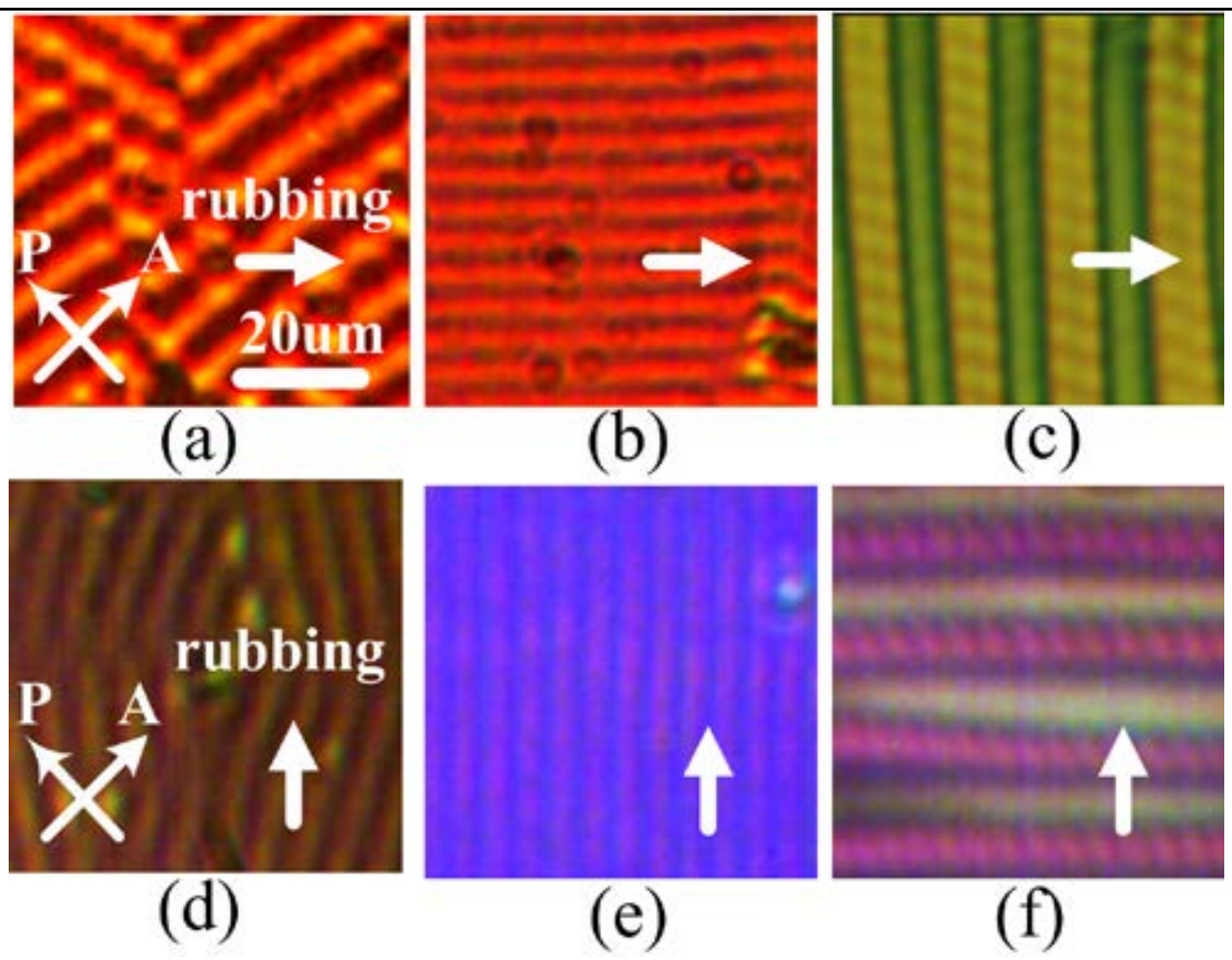

**FIG. 4** (Color online) Characteristic pattern morphologies observed in a polarizing microscope: (a) and (d) oblique stripes with ≈ 9μm periodicity at $T$-$T_c$= -5$^o$C, $f$ = 80Hz, $V$ = 40$V_{pp}$; (b) and (e) longitudinal stripes with ≈5 μm periodicity at $T$-$T_c$= -5$^o$C, $f$ = 500Hz, $V$ = 80$V_{pp}$; (c) and (f) normal stripes with ≈ 17 μm periodicity at $T$-$T_c$= -1$^o$C, $f$ = 500Hz, $V$ = 160$V_{pp}$. The inscriptions A, P and rubbing in the microphotographs (a) and (d) refer to the analyzer, polarizer and the rubbing direction of the LC cell, respectively. The different colours in morphologies according to different orientation indicate the birefringence of these stripes.

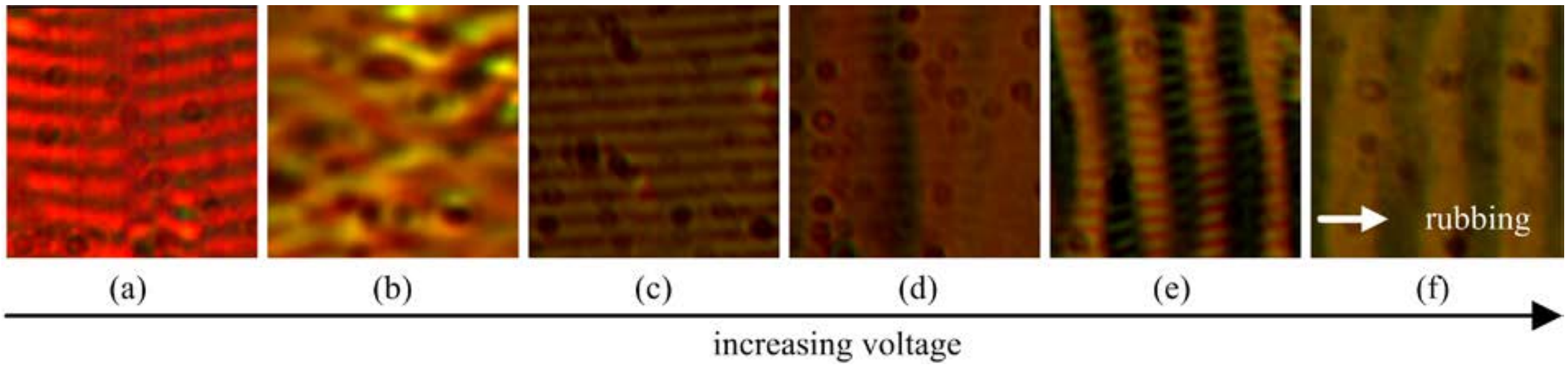

**FIG. 5** (Color online) Voltage dependence of the patterns at the temperature $T$-$T_c$ = -3 °C and frequency $f$ = 500 Hz, illustrated by a sequence of microphotographs: (a) oblique stripes at $V$ = 58 $V_{pp}$; (b) transformation from oblique to longitudinal stripes at $V$ = 62 $V_{pp}$; (c) longitudinal stripes at $V$ = 86 $V_{pp}$; (d) transformation from longitudinal to normal stripes at $V$ = 110 $V_{pp}$; (e) normal stripes (coexisting with the fine longitudinal ones) at $V$ = 146 $V_{pp}$; (f) normal stripes at $V$ = 250 $V_{pp}$ before entering the chaotic regime. The microphotographs cover an area of 50 μm × 50 μm. The arrow in (f) indicates the initial director orientation (rubbing direction).

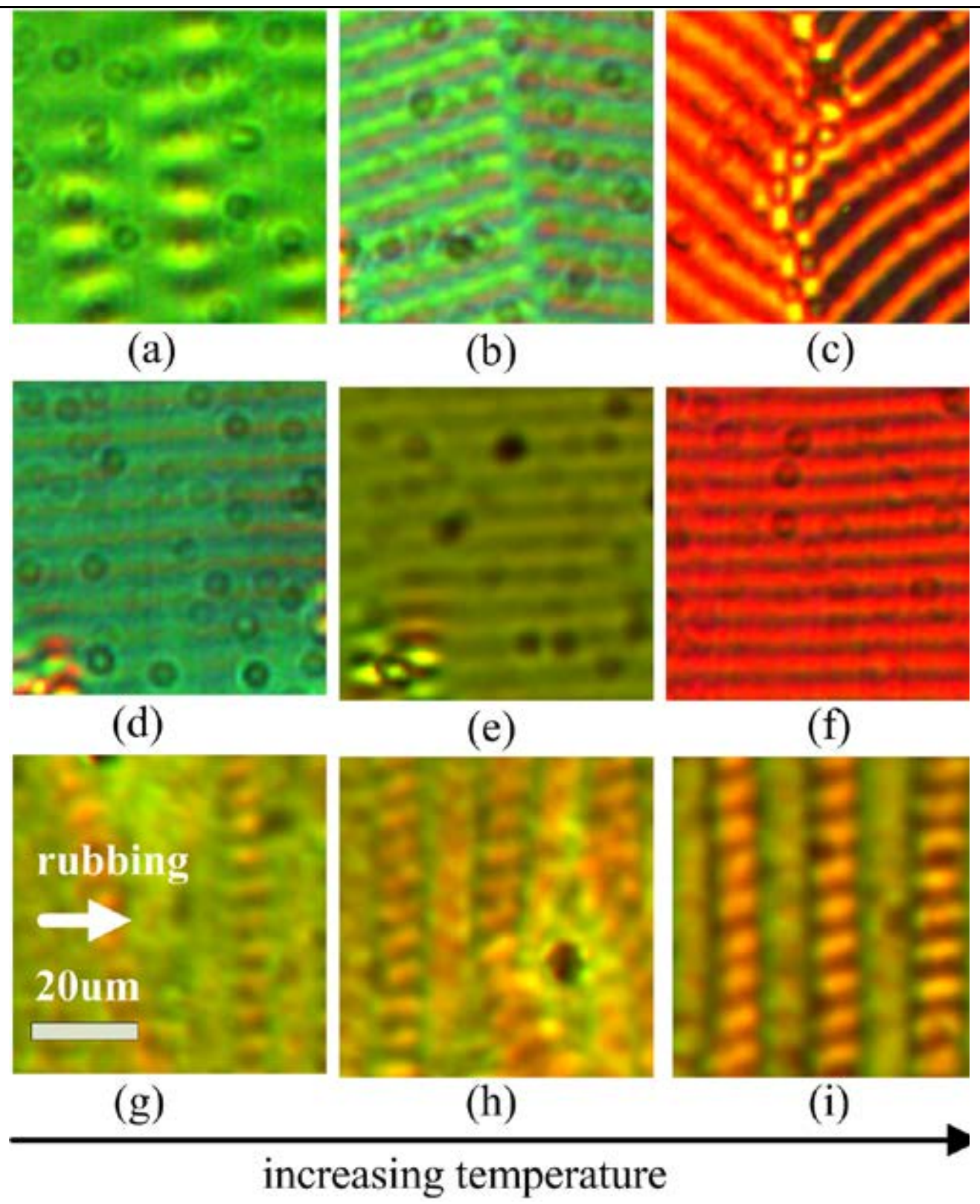

**FIG. 6** (Color online) The effect of temperature increase on various pattern morphologies: oblique stripes ($f$ = 125 Hz) at (a) $T$ - $T_c \approx$ -15 ℃, $V \approx$ 12 $V_{pp}$, (b) $T$ - $T_c \approx$ -10 ℃, $V \approx$ 32 $V_{pp}$, (c) $T$ - $T_c \approx$ -5 ℃, $V \approx$ 46 $V_{pp}$; longitudinal stripes ($f$ = 500 Hz) at (d) $T$ - $T_c \approx$ -15 ℃, $V \approx$ 70 $V_{pp}$. (e) $T$ - $T_c \approx$ -10 ℃, $V \approx$80 $V_{pp}$. (f) $T$ - $T_c \approx$ -5 ℃, $V \approx$ 84 $V_{pp}$; normal stripes ($f$ = 400 Hz) at (g) $T$ - $T_c \approx$ -12 ℃, $V \approx$ 120$V_{pp}$, (h) $T$ - $T_c \approx$ -10 ℃, $V \approx$ 120 $V_{pp}$, (i) $T$ - $T_c \approx$ -5 ℃, $V \approx$ 120 $V_{pp}$.

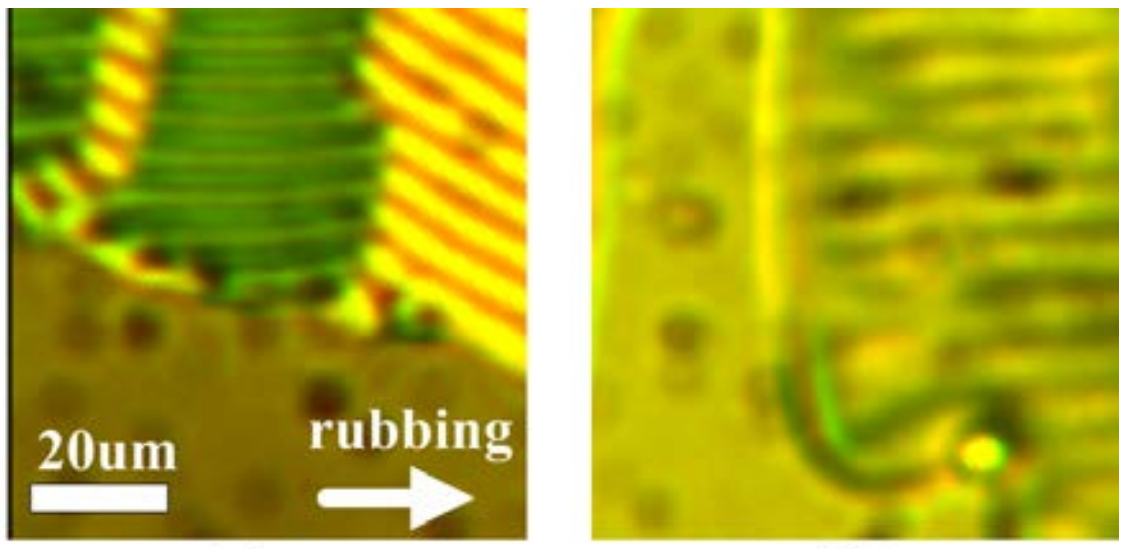

(a) (b)

**FIG. 7** (Color online) Electroconvection at the clearing temperature ($T = T_c$) . The microphotographs show a coexistence of the nematic (upper right corner) and isotropic (bottom left corner) phases separated by a sharp phase boundary line. EC pattern exists in the nematic part: (a) oblique stripes ($f$ = 150 Hz) at $V \approx 60\ V_{pp;}$ (b) longitudinal stripes ($f$ = 500 Hz) at $V \approx 88\ V_{pp}$.

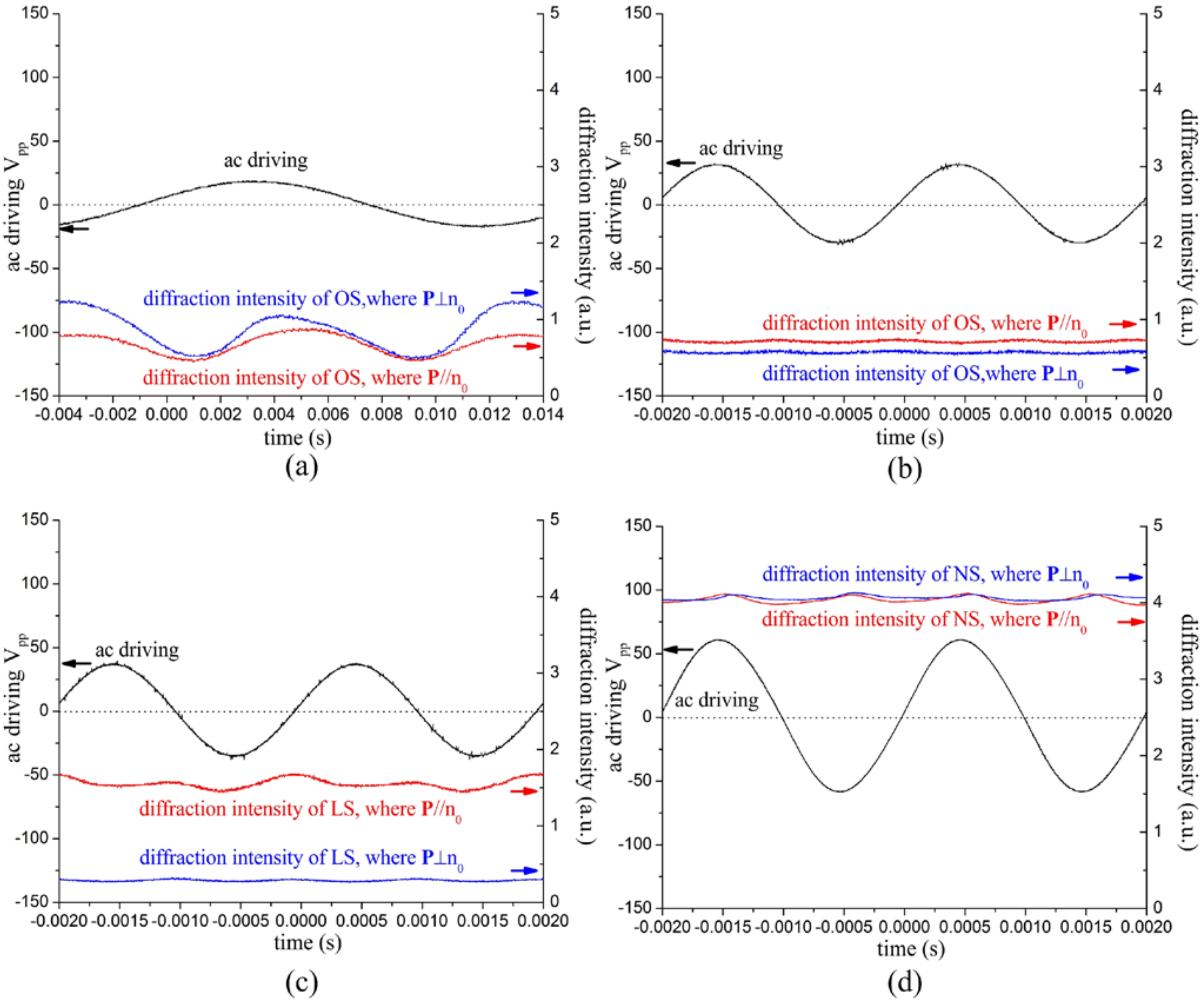

**FIG. 8** (Color online) Temporal evolution of the applied voltage and of the 1st order diffraction intensity at two orthogonal polarization directions of the illuminating light for various patterns under ac driving at the temperature $T$-$T_c \approx$ -3 ℃: (a) oblique stripes (OS) at $V \approx$ 35 $V_{pp}$, $f$ = 60 Hz; (b) oblique stripes (OS) at $V \approx$ 64 $V_{pp}$, $f$ = 500 Hz; (c) longitudinal stripes (LS) at $V \approx$ 76 $V_{pp}$, $f$ = 500 Hz; (d) normal stripes (NS) at $V \approx$ 120 $V_{pp}$, $f$ = 500 Hz.

**Table.1** The temperature dependence of the onset voltages $V_c$ at $f$ = 500 Hz in the vicinity of the clearing point $T_c$.

| Temperature deviation ($T$-$T_c$) | oblique $V_{cOS}$ | longitudinal $V_{cLS}$ | normal $V_{cNS}$ |
|---|---|---|---|
| -0.5 ℃ | 57 $V_{pp}$ | 73 $V_{pp}$ | 123 $V_{pp}$ |
| -1.0 ℃ | 53 $V_{pp}$ | 71 $V_{pp}$ | 120 $V_{pp}$ |
| -2.0 ℃ | 51 $V_{pp}$ | 69 $V_{pp}$ | 118 $V_{pp}$ |
| -3.0 ℃ | 48 $V_{pp}$ | 67 $V_{pp}$ | 116 $V_{pp}$ |
| -4.0 ℃ | 45 $V_{pp}$ | 64 $V_{pp}$ | 110 $V_{pp}$ |